\documentclass[runningheads]{llncs}

\usepackage[utf8]{inputenc}
\usepackage{zref}
\usepackage{biblatex}
\usepackage{subcaption}
\usepackage{graphicx}
\usepackage{comment}
\usepackage[hidelinks]{hyperref}
\usepackage[noabbrev,nameinlink]{cleveref}

\usepackage{listings}
\usepackage{multirow}
\usepackage{xcolor}
\definecolor{mygray}{gray}{0.96}
\definecolor{mypurple}{rgb}{0.6, 0, 0.6}

\lstdefinestyle{mystyle}{
        language=C,
        frame=single,
        numbers=left,
        tabsize=2,
        numberstyle=\tiny,
        numbersep=5pt,
        backgroundcolor=\color{mygray},
        captionpos=b, 
        keywordstyle=\color[rgb]{0, 0.6, 0},
        commentstyle=\color[rgb]{0,0,1},
        stringstyle=\color{red},
        morecomment=[l][\color{mypurple}]{\#},
        basicstyle=\scriptsize,
        showtabs=false,
        showspaces=false,
        showstringspaces=false,
        breaklines=true,
        escapeinside={(*@}{@*)},
        captionpos=b,
}

\title{On the Effect of Clock Frequency on Voltage and Electromagnetic Fault Injection}
\subtitle{}
\author{Stefanos Koffas\inst{1} \and Praveen Kumar Vadnala\inst{2}}
\institute{Delft University of Technology, Delft, The Netherlands, \email{s.koffas@tudelft.nl} \and
Riscure BV, Delft, The Netherlands, \email{vadnala@riscure.com}}

\begin{document}
\maketitle

\begin{abstract}

We investigate the influence of clock frequency on the success rate of a fault injection attack. In particular, we examine the success rate of voltage and electromagnetic fault attacks for varying clock frequencies. Using three different tests that cover different components of a System-on-Chip, we perform fault injection while its CPU operates at different clock frequencies. Our results show that the attack's success rate increases with an increase in clock frequency for both voltage and EM fault injection attacks. As the technology advances push the clock frequency further, these results can help assess the impact of fault injection attacks more accurately and develop appropriate countermeasures to address them.

\end{abstract}


\section{Introduction}
\label{sec:intro}

Fault Injection (FI) attacks have been used to attack cryptographic implementations for over two decades. It is now well known that both symmetric and asymmetric cryptosystems are vulnerable to Differential Fault Analysis (DFA) attacks~\cite{Boneh97onthe, biham1997differential, giraud2004dfa, biehl2000differential}. However, breaking cryptographic implementations is just one of the many possibilities for FI attacks. They have been frequently used to break the security of smart cards and embedded devices~\cite{vfi-shaping-the-glitch, escalating-privileges, controling-pc, vasselle2018laser}.
FI attacks have been successfully used to break secure boot, e.g., bypassing the authentication of the code stored in flash memory, allowing attackers to run their code on the device. Further, FI has been used for privilege escalation or to extract firmware from the device.

\emph{\textbf{Previous Work.}} Boneh, DeMillo, and Lipton demonstrated how faults induced in hardware could be exploited to recover the secret key used in RSA~\cite{Boneh97onthe}. In this attack, a fault is injected while the device performs an RSA operation, leading to incorrect output. Given several incorrect outputs and the correct output, recovering the secret key used with a DFA attack is possible. Similar attacks have been later proposed for other public, and symmetric-key algorithms~\cite{giraud2004dfa, biham1997differential, biehl2000differential, aumuller2002fault, anderson1997low}. A survey of these successful fault attacks can be found in~\cite{FI_crypto_survey}.

Moreover, existing published results use FI to break the non-cryptographic security mechanisms. In~\cite{controling-pc}, Timmers, Spruyt, and Witteman showed that FI could be used to load attacker-controlled data into the Program Counter (PC) register in an ARM 32-bit platform, allowing an attacker to gain runtime control of the device by setting the PC to an address where the attacker's payload is stored. In~\cite{riviere2015high}, the authors performed FI on the instruction cache of ARMv7-M architectures and modified the control flow of a program.
Cui and Housley used FI to corrupt the data stored in DRAM, thereby breaking the secure boot of an embedded device~\cite{cui2017badfet}.
A laser FI attack has been successfully used to break the secure boot of a smartphone in~\cite{vasselle2018laser}. In~\cite{escalating-privileges}, FI has been successfully used to escalate the privileges in Linux from user mode to kernel mode. Recently, FI has also been used to extract the firmware from several commercial devices~\cite{vfi-shaping-the-glitch}.

\emph{\textbf{Contributions.}} Unlike smart cards, many embedded systems in use today are implemented using multi-core System-on-Chips (SoCs) that are complex and host CPUs that run at hundreds of MHz to few GHz. Most of these SoCs can operate at different frequencies, and they often provide an option to configure their frequency externally or internally. Moreover, some SoCs start booting directly from an external clock that is relatively slow and switch to PLL (Phase Locked Loop) sometime during the boot flow. This switch leads to a natural question: \textit{does the FI's attack success rate depend on the operating frequency?}

The success rate of an FI test is defined as the number of successful faults divided by the number of total attempts. So, naturally, as the success rate increases, the effort required to perform a successful attack decreases. Although dependency between the EM pulse voltage and the clock frequency along with success rate was briefly discussed in~\cite{ordas2014evidence} within the context of FPGA, to the best of our knowledge, no extensive study examined the relationship between the clock frequency and the success rate of an FI attack. In this work, we address this gap for Voltage FI (VFI) and ElectroMagnetic FI (EMFI) within the context of an SoC. We use SiFive's HiFive1 development board for our experiments, which houses the FE310-G000 chip, the first commercially available RISC-V SoC.

\emph{\textbf{Organization.}} The rest of the paper is organized as follows. We provide a brief introduction to different FI attacks and fault models in~\autoref{sec:prel}. Next, we describe the three different test applications used in our testing in~\autoref{sec:tests}. The hardware and the software tools used for the experiments are listed in~\autoref{sec:setup}. The results from our experiments for both VFI and EMFI on HiFive1 are presented in~\autoref{sec:exp}. We provide possible reasons for the observed behaviour in~\autoref{sec:disc}. Finally, we conclude the paper in~\autoref{sec:conc}.

\section{Preliminaries}
\label{sec:prel}

Fault injection attacks are a class of physical attacks that try to actively modify the intended behavior of the device in order to bypass its security. Faults can be injected into the targeted device through different means, e.g., varying the supply voltage or the clock speed, or using electromagnetic emissions or laser beams~\cite{bar2006sorcerer}. In this section, we describe the common techniques used to inject faults. We also recall various fault models from the literature.

\emph{\textbf{Clock Fault Injection.}} A fault is injected by tampering with the target's clock signal~\cite{balasch2011depth}. For example, the target is supplied with a clock signal higher than its operating frequency for a short period reducing the length of a single clock cycle.  Thus, it may cause setup time constraint violations~\cite{zussa2013power} changing the program's control flow, which could result in breaking a security mechanism.

\emph{\textbf{Voltage Fault Injection}} A fault is injected by changing the target's supply voltage~\cite{zussa2013power}. This change is applied when the targeted operation is executed, making it possible to induce the desired effect in the device. As shown in~\cite{zussa2013power}, voltage fault injection causes setup time violations like the clock fault injection.

\emph{\textbf{Electromagnetic Fault Injection}} A fault is injected by applying a transient or a harmonic EM pulse~\cite{dehbaoui2012electromagnetic, bayon2012contactless, moro2013electromagnetic}. A fault injection probe consisting of a coil generates such pulses after a high voltage pulse is applied to the coil, inducing eddy currents into the chip. These eddy currents cause faulty behavior that could be used to break a security mechanism.

\emph{\textbf{Optical Fault Injection.}} A fault is injected into the target device with the help of a light pulse~\cite{skorobogatov2002optical}. The applied light pulse induces a photo-electric current in the device, causing faults in the computations. The light pulse can be generated using a low-cost camera flashlight, but often this is not precise. For higher precision, a laser beam is used to induce the desired light pulse.

\emph{\textbf{Fault Models.}}
The behavior of a device can be affected in various ways due to fault injection attacks. These attacks can influence both the CPU's execution unit and the static components that store data and instructions like the registers and the caches~\cite{escalating-privileges}. In general, it is difficult to determine the exact reason behind a successful fault injection attack. Therefore, we use high-level fault models that describe the effect of faults on the device's behavior on the instruction set architecture level~\cite{controling-pc}. Commonly used fault models include:

\begin{itemize}
	\item \textbf{Instruction Manipulation}: The fault modifies the instruction, leading to unexpected behavior. For example, a bit flip in the opcode field of an instruction converts a subtraction operation into an addition.
	\item \textbf{Instruction Skipping}: This is a special case of instruction manipulation that results in a modified instruction that has no impact on the device's behavior. This can happen, for example, when the operands of the modified instruction have been changed to something that is not used later by the program or when a branch instruction has been changed to a \emph{nop}, i.e., no operation.
	\item \textbf{Memory Corruption}: The fault affects the values loaded from a register or memory, which can cause unexpected effects on the program execution. This can happen when the data loaded to a register from the data cache is corrupted. Alternatively, when the data read from the main memory is corrupted, the data or instruction cache stores the corrupted value.
\end{itemize}

\section{Test Applications}
\label{sec:tests}

In this section, we propose three test applications that aim to capture the effects of faults on different SoC components. These tests are based on the characterization test presented in~\cite{controling-pc} and intend to cover the effect of faults on an SoC more extensively. At a high level, a fault can modify the instructions being executed or the data being processed through a single or multiple bit flips. Such modifications can occur in any SoC component, like the CPU or the memory, or during data exchange. We aim to cover different scenarios where a fault could modify the data or instructions.

Our tests are designed to cover the effects on various SoC components. We implemented them in assembly to fully control what is being executed on the CPU and avoid any undesired effects caused by compiler optimizations. We show these tests in~\autoref{lst:test1},~\autoref{lst:test2}, and~\autoref{lst:test3} in an assembly-like pseudo-code that can be easily translated into any Instruction Set Architecture (ISA). In all these tests, we use two general-purpose registers named $t0$ and $t1$. Their names come from the temporary registers defined in RISC-V but all ISAs have such registers.

\subsection{Register-based Loop}
In the register-based loop, we only use the CPU registers to implement a loop. We use two counters: one that goes up and the other goes down. These counters are initialized to 0 ($t0$ register) and \emph{n} ($t1$ register), respectively. The test consists of a loop that increments and decrements $t0$ and $t1$, respectively, in steps of 1, until $t1$ becomes 0. The rest of the registers are initialized with a known fixed value (e.g., 0xdeadbeef) to monitor if the fault modified the source or destination registers in an instruction.
The test uses only two registers to store the counters, and hence the data cache will not be used. Additionally, as the code size is small, it should most likely fit in the instruction cache. 

A successful fault is identified by checking the value of the registers at the end of the loop. In some cases, the registers $t0$ and $t1$ do not hold the values \emph{n} and zero due to the injected fault. Alternatively, the fault could also affect the value in the unused registers.

\begin{lstlisting}[caption=Register based test,
                   escapeinside={(*}{*)},
                   label=lst:test1]
	# Push a known value to all the registers
	# N: the number of registers in the ISA
	(t2, ..., tN) (*$\leftarrow$*) 0xdeadbeef
	t0 (*$\leftarrow$*) 0
	t1 (*$\leftarrow$*) n
	reg_loop:
	   t0 (*$\leftarrow$*) t0 (*\texttt{+}*) 1
	   t1 (*$\leftarrow$*) t1 (*\texttt{-}*) 1
	   if t1 (*\texttt{>}*) 0 then goto reg_loop
\end{lstlisting}

\subsection{Memory-based Loop}
The memory-based loop is similar to the register-based loop but the counters are loaded/saved from/to the memory (using \texttt{load} and \texttt{store} instructions) in every iteration. Again, the loop ends when $t1$ is 0 (\autoref{lst:test2}). After the first load, a copy of the data is kept in the data cache, and hence faults would only affect the data cache and its transfers inside the loop. The loop code should fit in the instruction cache due to its size. We also initialize all the unused registers to a fixed value to track any corruptions in their contents or verify whether a different register was used in a loop iteration due to the fault. A successful fault is determined by examining the registers and comparing their values with the expected ones.

\begin{lstlisting}[caption=Memory based test,
                   escapeinside={(*}{*)},
                   label=lst:test2]
	# Push a known value to all the registers
	# N: the number of registers in the ISA
	(t2, ..., tN) (*$\leftarrow$*) 0xdeadbeef
	t0 (*$\leftarrow$*) 0
	stack[sp (*\texttt{-}*) 4] (*$\leftarrow$*) t0
	t1 (*$\leftarrow$*) n
	stack[sp (*\texttt{-}*) 8] (*$\leftarrow$*) t1
	mem_loop:
	   t0 (*$\leftarrow$*) stack[sp (*\texttt{-}*) 4]
	   t1 (*$\leftarrow$*) stack[sp (*\texttt{-}*) 8]
	   t0 (*$\leftarrow$*) t0 (*\texttt{+}*) 1
	   t1 (*$\leftarrow$*) t1 (*\texttt{-}*) 1
	   stack[sp (*\texttt{-}*) 4] (*$\leftarrow$*) t0
	   stack[sp (*\texttt{-}*) 8] (*$\leftarrow$*) t1
	   if t1 (*\texttt{>}*) 0 then goto mem_loop
\end{lstlisting}

\subsection{Unrolled Loop}
In this test, we implement a fully unrolled loop. We use one up-counter ($t0$) initialized to 0, that is incremented \emph{n} times through an unrolled loop. Similar to the other two tests, we also initialize all the unused registers to a fixed value.

In general, this test can be used in two different ways according to the loop's number of increment instructions. First, if a small \small{n} is used, the program can fully fit in the instruction cache, which results in no cache misses during the execution of the test. As a result, only transfers between the instruction cache and the CPU are affected. This way, it is possible to pinpoint the sensitivity of the instruction cache and the corresponding bus to FI attacks. On the other hand, if a large \emph{n} is used and the program cannot fit in the instruction cache, there will be instruction cache misses during the execution of the test, which results in loading the instructions from the main memory. The CPU to main memory bus's sensitivity to FI attacks could also be determined in such cases.

\begin{lstlisting}[caption=Unrolled loop,
                   escapeinside={(*}{*)},
                   label=lst:test3]
	# Push a known value to all the registers
	# N: the number of registers in the ISA
	(t1, ..., tN) (*$\leftarrow$*) 0xdeadbeef
	t0 (*$\leftarrow$*) 0
	t0 (*$\leftarrow$*) t0 (*\texttt{+}*) 1
	t0 (*$\leftarrow$*) t0 (*\texttt{+}*) 1
	t0 (*$\leftarrow$*) t0 (*\texttt{+}*) 1
	...
	t0 (*$\leftarrow$*) t0 (*\texttt{+}*) 1
\end{lstlisting}

A successful fault is detected when the value in $t0$ is not equal to \emph{n} or when the value in any of the unused registers is corrupted.

\section{Setup}
\label{sec:setup}

Performing automatic execution of FI attacks requires both hardware and software tools. In this section, we describe the tools used for our experiments. We also briefly discuss the characteristics of the target used for the experiments.

\subsection{Target of Evaluation}

Our target is the FE310-G000 (fabricated in TSMC CL018G 180nm~\cite{fe310-g000-manual}) which is included in the development board HiFive1.
FE310-G000's maximum supported frequency is 320MHz and the CPU requires 1.8V or 3.3V supply voltage to operate~\cite{fe310-g000-datasheet}. We did not decap the chip due to its thin package as was also described in~\cite{emfi-on-arm-and-riscv}.
The tests presented in~\autoref{sec:tests} are implemented as part of a user-defined program that runs on bare-metal (without an operating system), as described in~\autoref{sec:test-app}.

\subsection{Hardware Tools}
During our experiments, various faults (also referred to as attempts in this paper) are injected into the device in order to identify suitable parameters for success. For that reason, a fully automated setup has been created using commercially available tools from Riscure~\cite{riscure_tools}. We used the following hardware tools for testing:

\begin{itemize}
	\item \textbf{Glitch Generator:} An FPGA based workbench that can be programmed to interact with embedded devices. The ``brain" of this device consists of two finite state machines (up to 255 states each), which are responsible for the correct generation of every signal that is needed for our experiments. To handle inputs/outputs, the device consists of 32 GPIO pins that can interact with the target. We use this device to generate the glitch used for the fault injection. The Glitch Generator consists of six analog voltage outputs, and it can also provide the input voltage for small embedded devices.
	\item \textbf{Glitch Amplifier:} This analog device is used in conjunction with the Glitch Generator. It is used to generate sharper and more accurate voltage glitches that are essential in voltage fault injection attacks.
	\item \textbf{EMFI Transient Probe:} The Glitch Generator controls this device. It generates an electromagnetic pulse lasting for 50ns after the Glitch Generator triggers it. The probe is made of a copper winding around a ferrite core, and its tip is a flat circle with a diameter of 1.5mm.
\end{itemize}

\subsection{Software Tools}
\label{sec:test-app}
We implemented a simple bare-metal application in C and RISC-V assembly that runs on the FE310-G000 chip. This application accepts messages from the PC through the UART interface and runs one of the three characterization tests described in~\autoref{sec:tests}. The message from the UART determines which test should run every time. Right before the test starts, a GPIO pin is set to high, and the same pin is set to low when the test ends. This pin is used in the synchronization between the Glitch Generator and the target device. The application can alter the chip’s operating frequency dynamically by enabling or disabling the PLL. We used three clock configurations to investigate the effect of different operating frequencies on both VFI's and EMFI's attack success rate. The first configuration operates at 16MHz (\emph{slow} configuration) and does not use the PLL clock generator. However, the \emph{medium} and the \emph{fast} configurations use the PLL. The \emph{medium} configuration operates at 90MHz. The \emph{fast} configuration operates at 320MHz and 240MHz for the EMFI and the VFI, respectively. We had to operate the device slightly slower than the maximum allowed frequency for VFI due to the instabilities introduced after the board was modified (see~\autoref{sec:vfi-setup}). In~\autoref{tab:commands}, we show a summary of the protocol used between the PC and the board.

\begin{table}[ht]
    \centering
    \caption{The list of commands used to communicate with the board}
	\resizebox{0.6\textwidth}{!}{%
	\begin{tabular}{|c|c|}
    \hline
        \textbf{Command} & \textbf{Functionality} \\ \hline
        "\texttt{\#}1"      & Run the register based loop (test 1) \\ \hline
        "\texttt{\#}2"      & Run the memory based loop (test 2) \\ \hline
        "\texttt{\#}3"      & Run the unrolled loop (test 3) \\ \hline
        "\texttt{\#}4"      & Enable the PLL at 320MHz (fast EMFI configuration) \\ \hline
        "\texttt{\#}5"      & Disable the PLL and use 16MHz (slow configuration) \\ \hline
        "\texttt{\#}6"      & Enable the PLL at 90MHz (medium configuration)  \\ \hline
        "\texttt{\#}7"      & Enable the PLL at 240MHz (fast VFI configuration)   \\ \hline
    \end{tabular}
	}
    \label{tab:commands}
\end{table}

\subsection{EMFI Setup}
\begin{figure}[!ht]
	\centering
	\resizebox{\textwidth}{!}{%
	\begin{subfigure}{0.48\textwidth}
		\centering
		\includegraphics[width=\linewidth]{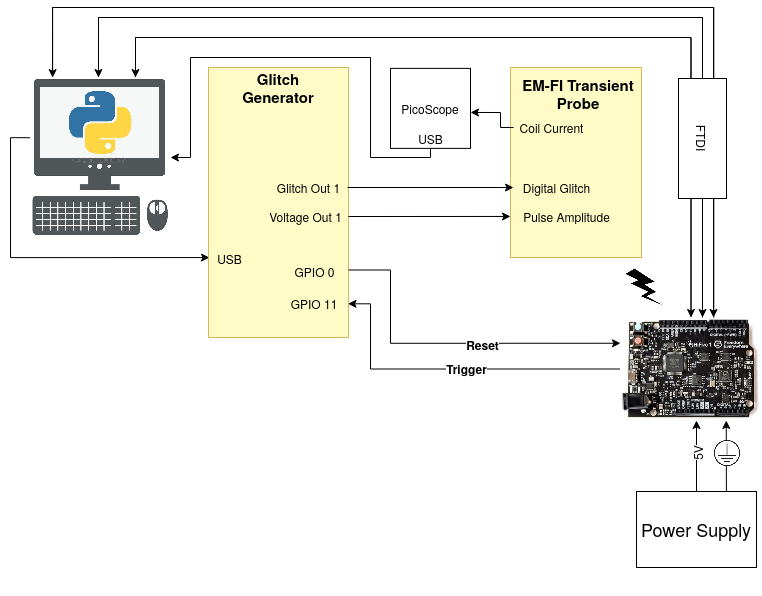}
		\caption{EMFI Setup}
		\label{fig:emfi-setup}
	\end{subfigure}
	\hfill
	\begin{subfigure}{0.48\textwidth}
		\centering
		\includegraphics[width=\linewidth]{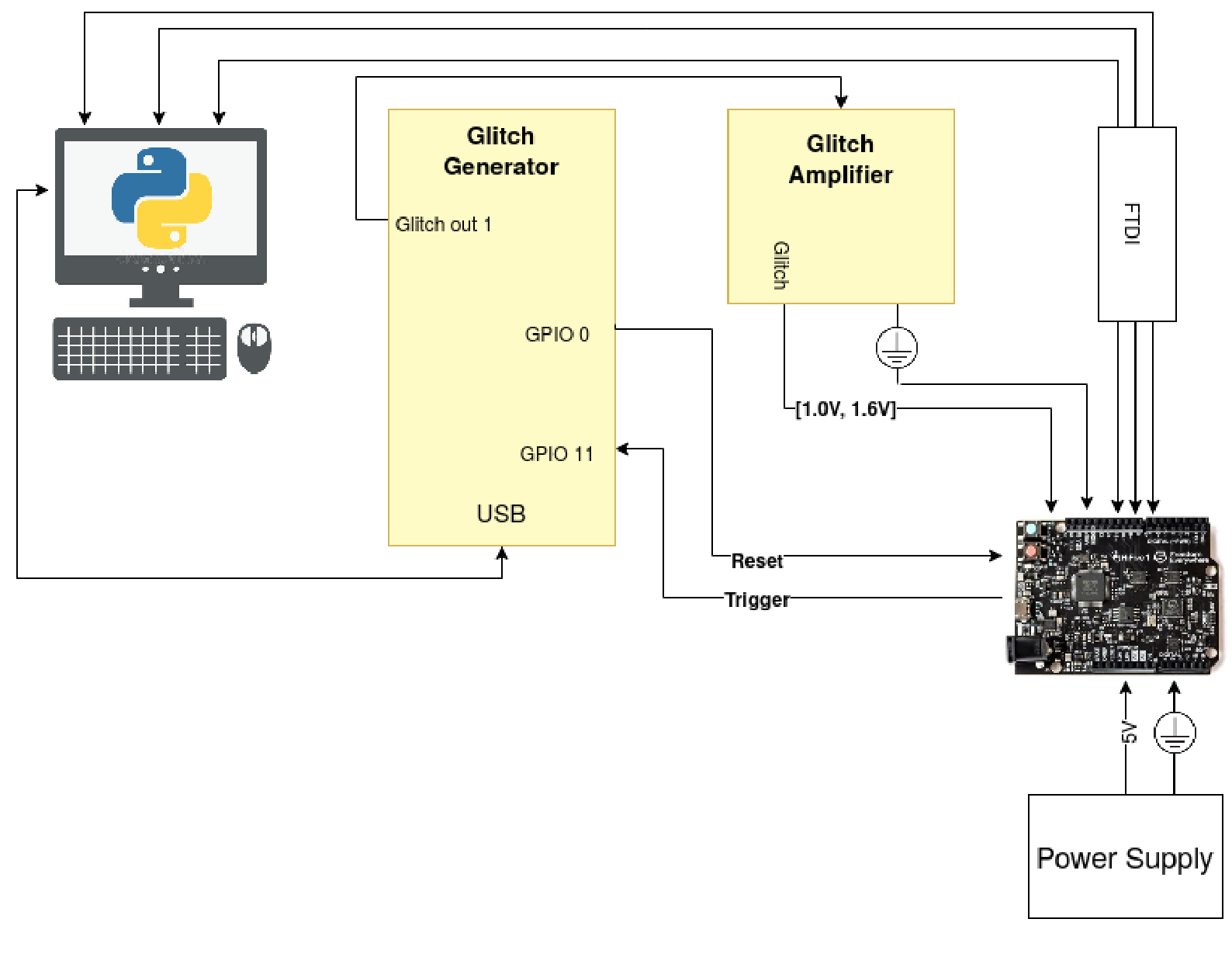}
		\caption{VFI setup}
		\label{fig:vfi-setup}
	\end{subfigure}
	}
	\caption{Setups used for the experiments.}
	\label{fig:setup}
\end{figure}

Our experimental setup used for the EMFI is shown in~\autoref{fig:emfi-setup}. The target board is powered using an external power supply. A Python script that runs in the PC controls the target through the UART interface and configures the state machine inside the Glitch Generator through a user-friendly API. This state machine consists of one state that produces the glitch when the trigger is generated from the target device. We program the target with our test application as we described in~\autoref{sec:test-app}. A jumper wire drives the trigger signal from the target's GPIO to the Glitch Generator. The Glitch Generator produces another trigger that is driven to the EMFI Transient Probe. The EMFI Transient Probe generates an EM pulse, which may or may not affect the target. The Transient Probe is attached to a CNC (Computer Numerical Control) machine that acts as a movable XYZ stage helping in accurate positioning above the target device. After the application finishes its execution, the device replies back to the PC, and the results are saved to an SQLite database. If no reply has been received after a specific amount of time, the target is reset by the Glitch Generator using the target's reset pin. We used an FTDI chip for the communication between the PC and the target.

\subsection{VFI Setup}
\label{sec:vfi-setup}

\begin{figure}[!ht]
	\centering
	\resizebox{\textwidth}{!}{%
	\begin{subfigure}{0.46\textwidth}
		\centering
		\includegraphics[width=0.8\linewidth]{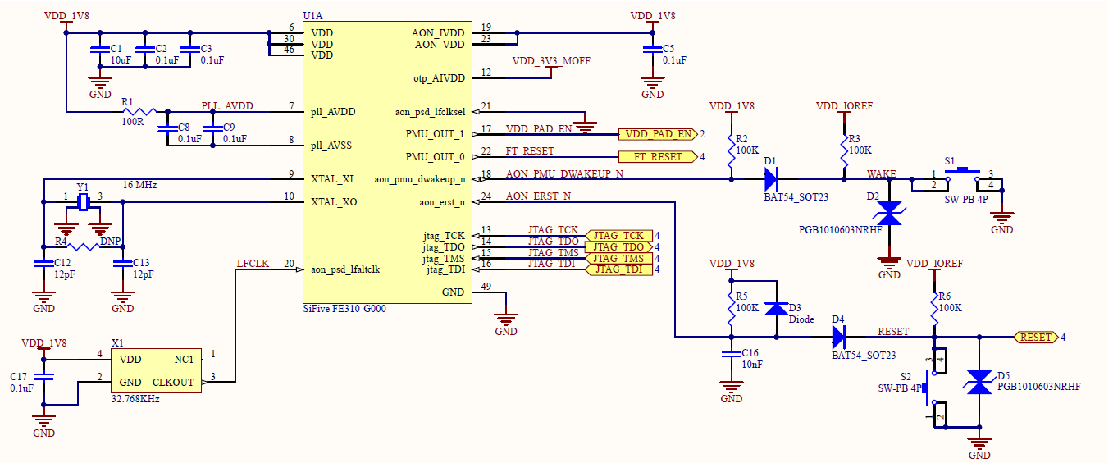}
		\caption{HiFive1 CPU schematic \cite{hifive1-schematics}}
		\label{fig:hifive1-cpu-schematics}
	\end{subfigure}
	\hfill
	\begin{subfigure}{0.37\textwidth}
		\centering
		\includegraphics[width=0.75\linewidth]{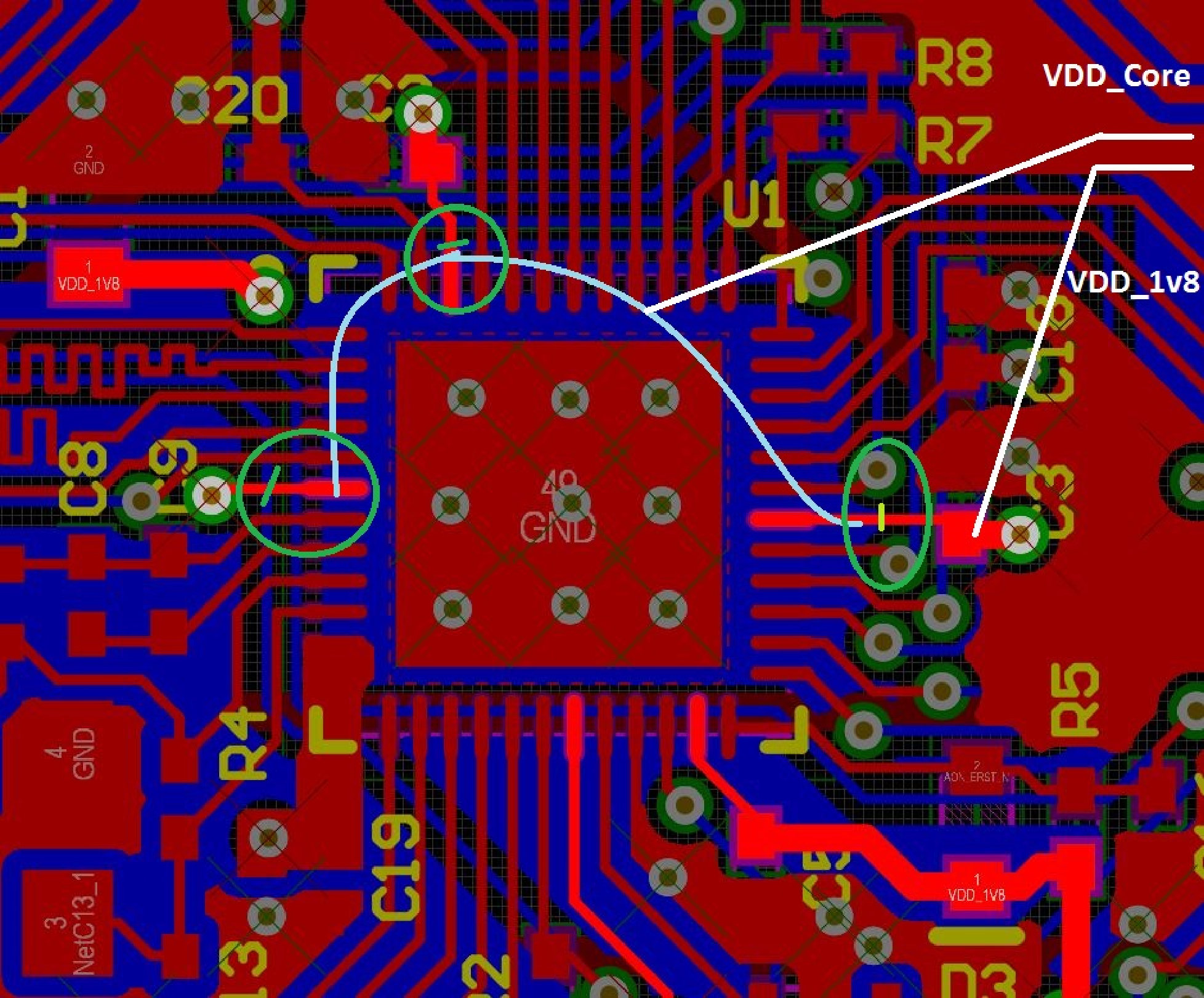}
		\caption{Power cuts in HiFive1 for effective voltage fault injection}
		\label{fig:hifive1-power-cuts}
	\end{subfigure}
	}
	\caption{HiFive1 CPU schematic and applied modifications.}
	\label{fig:schematic-and-power-cuts}
\end{figure}

In~\autoref{fig:hifive1-cpu-schematics}, the schematic that describes the circuit around the FE310-G000 chip in the HiFive1 development board is shown. We see that pins 6, 30, and 46 are used for the CPU's power supply. To create a stable power supply line that is not affected by small variations in the input voltage, some filter capacitors are connected directly to these pins. To increase the effectiveness of the VFI, the glitch should be applied directly to the CPU without having to pass through the filter circuit. Therefore, we modified the HiFive1 board used for our experiments. The applied power cuts are shown in~\autoref{fig:hifive1-power-cuts}. In particular, pins 6, 30, and 46 are cut from the rest of the circuit and soldered to an external pin so that it is possible to connect them to a 1.8V power supply directly (\texttt{VDD\_Core} in the \autoref{fig:hifive1-power-cuts}).

The experimental setup we used for VFI is shown in~\autoref{fig:vfi-setup}. For this experiment, the development board was powered from an external power supply at 5V. The power cuts isolated the SoC from the rest of the board, and hence we powered it separately. In particular, we connected the Glitch Amplifier directly to the external pin that powers it. When there is no glitch, the output of the Glitch Amplifier is set at 1.8V, as suggested in~\cite{fe310-g000-datasheet}. We verified that the chip correctly operated in this setup even though the filter capacitors have been removed. Like the EMFI, the target generates a trigger when the application starts. The trigger is then driven to the Glitch Generator. Next, the Glitch Generator produces the glitch, which is fed to the Glitch Amplifier. The glitch drops the $V_{in}$ to a random value smaller than 1.8V (from 1.0V to 1.6V) for a short time. We chose a broad range of values for the glitch to investigate the chip's behavior under various circumstances. When we need to perform an attack, we can narrow this range and use values that yield a high probability of a successful attack. A Python script controlled this process, and an FTDI chip was used for the communication between the PC and the target.

\subsection{Results Classification}
\label{sec:res-clas}

The output from a fault injection attempt is categorized as follows:

\begin{itemize}
	\item \textbf{Expected}: The test has completed its execution and the expected results (e.g., $t0$ = $n$ and $t1$ = $0$) were sent back to the PC.
	\item \textbf{Crash/Mute}: The impact of the glitch was strong, and the target crashed, or there was no reply from the target before the timeout expired. In this case, the glitch either affects the execution path and the application cannot continue or causes the target to reset.
	\item \textbf{Successful}: The counter values ($t0$ and $t1$) returned to the PC were different from the expected ($t0$ $\neq n$ or $t1$ $\neq 0$). Therefore, the injected fault produced an undesirable effect on the program execution without causing a crash.
\end{itemize}
Note that the success rate of an experiment is defined as the number of successful attempts divided by the total number of attempts.

\section{Experimental Results}
\label{sec:exp}

To investigate how clock frequency affects FI success rate, we performed VFI and EMFI experiments while the CPU was clocked at different frequencies. In this section, we present the results from these experiments.

\paragraph{Parameter Space.}
The glitch applied in every fault injection attempt is fully defined through a set of configurable parameters. These parameters form the parameter space for the experiment, and they are different for every type of FI attack.

\subsection{EMFI}
The effectiveness of the EMFI depends on the location of the probe. Thus, we need to identify the location that gives the maximum success rate. For that reason, we performed a scan over the $6\times 6$mm chip's package~\cite{fe310-g000-datasheet} in a two-dimensional grid of points. On every point in the grid, we performed multiple FI attempts for statistical analysis. In general, the grid's density (distance between different points) depends on the size of the chip and its package and the transient probe tip's area. When the probe tip is small, a dense grid can be used. On the other hand, when the probe tip is large, a sparse grid should be used. If the grid remains dense even with a large tip, every EMFI experiment can affect multiple points in the grid, introducing redundancy in the results. We used an $8\times 8$ grid of 64 points for our scan, which resulted in a step of 0.75mm. The diameter of our probe tip is 1.5mm, and such a step is reasonable to avoid interference between different grid points. The grid's origin (X $= 0$, Y $= 0$) corresponds to the lowest left corner of the chip's resin package (see FE310-G000 pinout in~\cite{fe310-g000-datasheet}). 

The parameter space for the EMFI consists of the following:
\begin{itemize}
	\item \textbf{Glitch power}: The EM pulse's amplitude as a percentage of the Transient Probe's maximum supported power. The maximum supported power corresponds to a 470V pulse. We saw that values above 80\% resulted in many resets and below 40\% seemed ineffective. For that reason, we used values between 40\% and 80\% of the maximum power.
	\item \textbf{Glitch delay}: The time between the trigger and the glitch. This should not be larger than the whole duration of the test that runs on the CPU. There was no need for exact timing in our experiments as they were loops and our aim was to draw a sensitivity map. Thus, the delay used was a random value between the 35\% and 65\% of each test's execution time. As expected, the exact ranges are different for every test and every clock frequency.
	\item \textbf{X}: The X coordinate (in micrometers) of the grid point.
	\item \textbf{Y}: The Y coordinate (in micrometers) of the grid point.
\end{itemize}

\begin{figure}[!ht]
	\centering
	\includegraphics[width=0.8\linewidth]{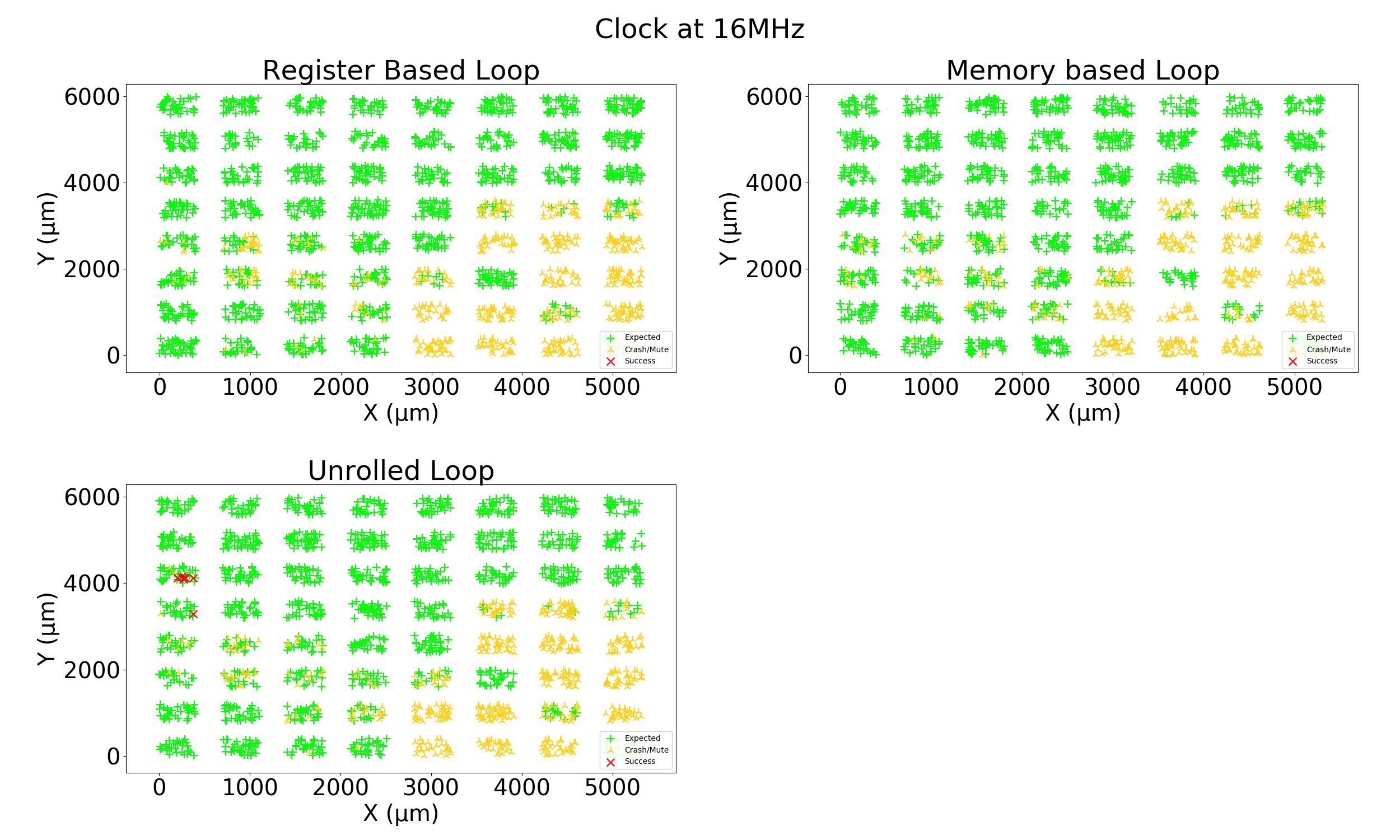}
	\caption{EMFI results of all three tests at 16MHz (PLL bypassed).}
	\label{fig:emfi-results-slow}
\end{figure}

\begin{figure}[!ht]
	\centering
	\includegraphics[width=0.8\linewidth]{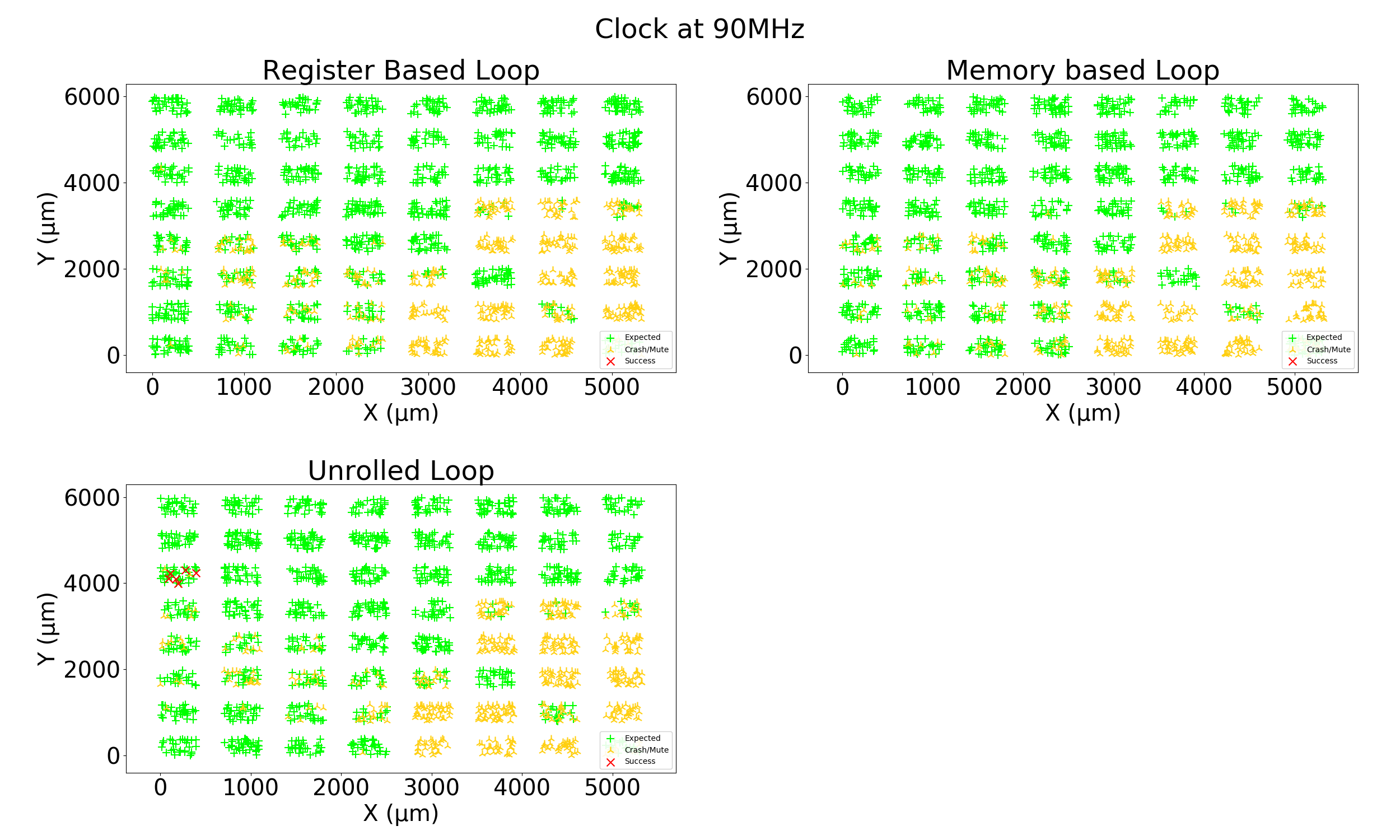}
	\caption{EMFI results of all three tests at 90MHz (PLL enabled).}
	\label{fig:emfi-results-medium}
\end{figure}

\begin{figure}[!ht]
	\centering
	\includegraphics[width=0.8\linewidth]{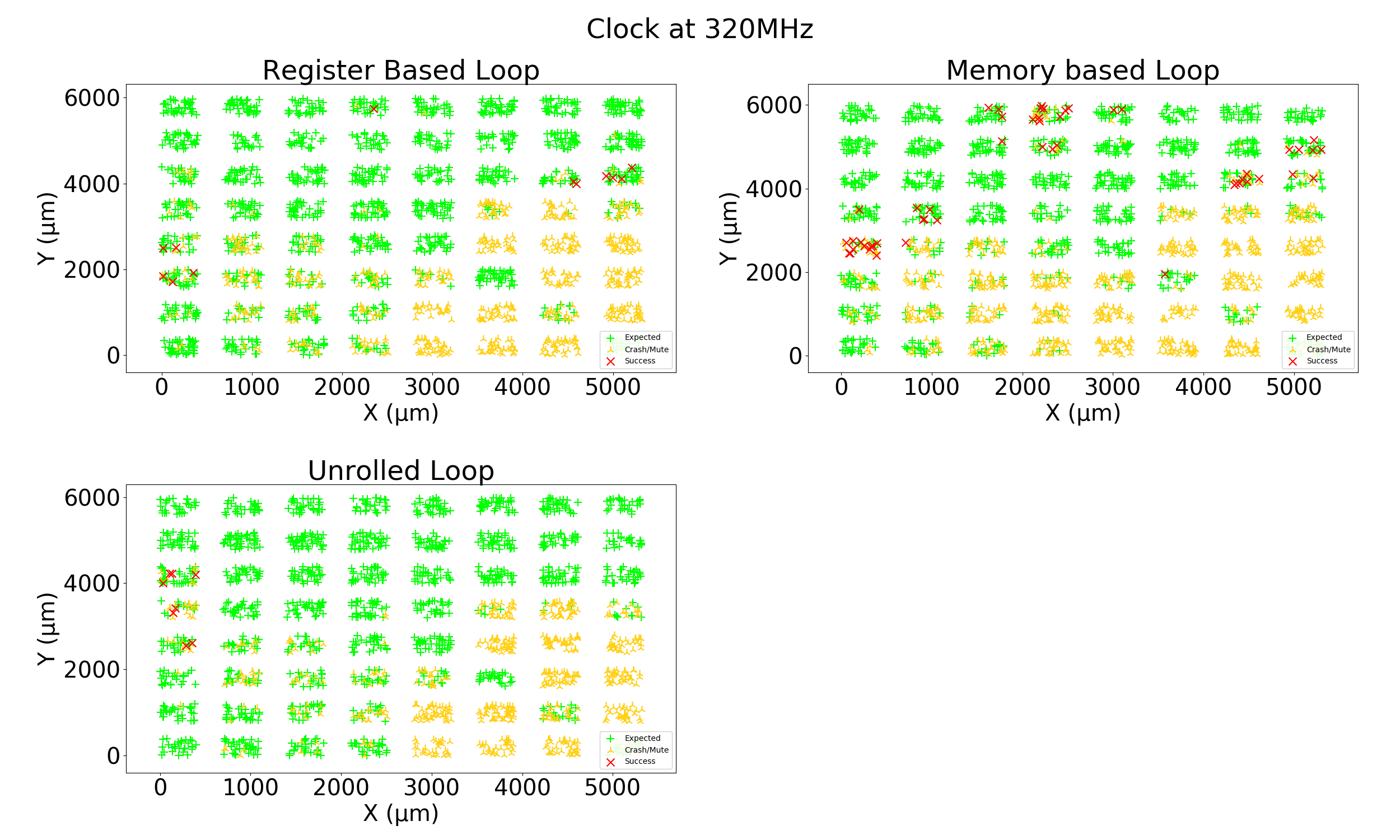}
	\caption{EMFI results of all three tests at 320MHz (PLL enabled).}
	\label{fig:emfi-results-fast}
\end{figure}

We scanned the whole chip package in this experiment. At each point, we performed a number of attempts with varying glitch delay and glitch power, both of which were selected randomly from the above predefined ranges. We set \emph{n} equal to 10000 (0x2710). The results of this experiment are summarized in~\autoref{fig:emfi-results-slow},~\autoref{fig:emfi-results-medium}, and~\autoref{fig:emfi-results-fast}. In these graphs, the x and y axes show the X and Y co-ordinates of the grid point, respectively. The \textit{green} color represents \textit{expected} results, the \textit{yellow} color \textit{crashes/mutes}, and the \textit{red} \textit{successful} results (see~\autoref{sec:res-clas}). We add a small random value (0-400 $\mu$m) to each experiment's X and Y coordinates so that they are not plotted on top of each other.

We see from~\autoref{fig:emfi-results-slow} and~\autoref{fig:emfi-results-medium} that for the slow (16MHz) and the medium (90MHz) clock configurations, successful glitches occurred only in the unrolled loop (\autoref{lst:test3}). These glitches occurred around the pins that communicate with the SPI flash memory (see F310-G000 pinout in~\cite{fe310-g000-datasheet}). The unrolled loop (\autoref{lst:test3}) consists of 10000 additions, and it requires an instruction cache of $10000 \: \textrm{instructions} * 4 \: 
\textrm{bytes per instruction} = 40000 \approx 39$KiB. However, FE310-G000 has only a 16KiB instruction cache. Thus, our faults were most likely affecting the instruction transfers from the flash.   
On the other hand, the code perfectly fits in the instruction cache when the number of loop iterations is small, i.e., 300. In this case, no successful glitch appeared around the region where we observed successful glitches previously. After the first iteration of the loop, the code is copied into the instruction cache, and there is no more interaction with the SPI flash memory. We saw no successful glitches in that case (slow and medium clock configurations), and we concluded that transfers from the instruction cache to the CPU were robust in these experiments. 

In~\autoref{tab:emfi-results-examples}, we summarize some of the observed successful attempts. For all the successful attempts in the unrolled loop test, the returned value is less than the expected 10000 (0x2710). If the returned counter value is close to 0x2710, it is safe to conclude that some \texttt{add} instructions were skipped due to the fault. This happened when $t0$ is 0x2700, 0x2708, and 0x270f for 16MHz, 90MHz, and 320MHz, respectively. In the remaining cases (i.e., 0x256e, 0x26c2, 0x2660), there is a significant difference between the returned and the expected values, so we cannot assume that only a few instructions were skipped. Even though the glitch lasts for 0.8, 4.5, and 16 cycles for the slow, medium, and fast clock configurations, the returned values differ by 418, 88, and 176 from the expected value. In~\cite{emfi-on-arm-and-riscv}, it was verified that an EMFI attack could affect multiple instructions at once. However, the probability of this effect dropped significantly for more than six instructions. Thus, in all these cases, some instructions were altered completely. Such alterations require multiple bit-flips for each altered instruction that is not aligned with the state-of-the-art fault models like the sampling~\cite{ordas2014evidence} or the charge-based~\cite{charge-based-fault-model}. Additionally, the successful attempts in this area were not increased as the operating frequency increased, which contradicts the charge-based fault model~\cite{charge-based-fault-model}. We concluded that this behavior was possible due to the retrieval of instructions from the external SPI flash. Therefore, these errors were not of particular interest for this work because the external SPI flash is unprotected in this board, and an attacker could directly attack it.

The other two tests have successful faults only when the chip operates at 320MHz, verifying the charge-based fault model~\cite{charge-based-fault-model}. In particular, we conclude that the branch instruction can be skipped in the register-based test, as one of the results was ($t0$, $t1$) = (7190, 2810) = (0x1c16, 0xafa). In that case, everything was run as expected until the branch instruction because $t0 + t1$= 10000. Additionally, in two cases a result larger than 10000 was saved in $t0$ (0x2e20 and 0x29ef), but the loop exited normally (i.e., $t1$ $= 0$). This indicates that the immediate values of the additions can be altered (instruction manipulation) because the significant difference from 10000 cannot be explained by a few instruction skips even if our glitch lasts for 16 clock cycles.

For the memory-based loop, we observed all possible faults when the chip operated at 320MHz. In the first case (($t0$, $t1$) = (0x270f, 0x0)), one addition was skipped. In the second case (($t0$, $t1$) = (0x62f, 0x20e1)), the branch instruction was skipped because $t0 + t1$ = 0x2710 and $t1$ $\neq 0$. The third case (($t0$, $t1$) = (0x31fe, 0)) shows that the constant added or subtracted has been manipulated and changed to a completely different value. Furthermore, memory corruption can also be seen when ($t0$, $t1$) = (0xdeadde040, 0). In this case, during the loop's execution, the contents of a register with the value 0xdeadbeef were saved in the stack. This value was retrieved in the next loop iteration. Then, the execution of the loop continued normally for 8529 (0xdeade040 - 0xdeadbeef) iterations until $t1$ was set to zero.
From~\autoref{fig:emfi-results-fast}, we infer that it is easier to induce successful faults when the target operates at the highest possible frequency. Among the three tests, the memory-based loop (\autoref{lst:test2}) has produced a higher percentage of successful faults because memory operations (i.e., loads and stores) are highly vulnerable to faults. Additionally, we practically verified that each program behaves differently under the same EMFI attacks highlighting the need for a profiling phase before targeting an application. Such a profiling phase could define the susceptibility of different assembly instructions to EMFI.

\begin{table}[]
	\centering
	\caption{Output from some of the successful faults in EMFI}
	\resizebox{0.85\textwidth}{!}{%
	\begin{tabular}{|c|c|c|c|c|}
	\hline
	Frequency & Test & t0 & t1 & Comment \\ \hline
	\multirow{4}{*}{16MHz} & register based loop & - & - & - \\ \cline{2-5}
	 & memory based loop & - & - & - \\ \cline{2-5}
		& \multirow{2}{*}{unrolled loop} & 0x256e & - & instruction manipulation (add) \\ \cline{3-5}
		&  & 0x2700 & - & instruction skipping (add) \\ \hline
	 \multirow{4}{*}{90MHz} & register based loop & - & - & - \\ \cline{2-5}
	 & memory based loop & - & - & - \\ \cline{2-5}
		& \multirow{2}{*}{unrolled loop} & 0x26c2 & - & instruction manipulation \\ \cline{3-5}
		&  & 0x2708 & - & instruction skipping (add) \\ \hline
	\multirow{9}{*}{320MHz} & \multirow{3}{*}{register based loop} & 0x1c16
		& 0xafa & instruction skipping  (branch) \\ \cline{3-5}
		&  & 0x2e20 & 0x0 & instruction manipulation (add) \\ \cline{3-5}
		&  & 0x29ef & 0x0 & instruction manipulation (add) \\ \cline{2-5}
	 & \multirow{4}{*}{memory based loop} & 0x270f & 0x0 & instruction
		skipping (add) \\ \cline{3-5}
		&  & 0x62f & 0x20e1 & instruction skipping (branch) \\ \cline{3-5}
		&  & 0x31fe & 0x0 & instruction manipulation (add) \\ \cline{3-5}
	 &  & 0xdeade040 & 0x0 & memory corruption \\ \cline{2-5}
		& \multirow{2}{*}{unrolled loop} & 0x2660 & - & instruction manipulation \\ \cline{3-5}
		&  & 0x270f & - & instruction skipping (add) \\ \hline
	\end{tabular}%
	}
	\label{tab:emfi-results-examples}
\end{table}

\subsection{VFI}
For an effective Voltage fault injection, we have removed the chip's filter capacitors (see~\autoref{sec:vfi-setup}). The CPU was powered directly from the Glitch Generator. Its supply voltage should be 1.8V (+/- 10\%) to function normally. For the glitch, the input voltage was dropped for a small amount of time, different for every clock configuration.

The parameter space for the VFI consists of the following:
\begin{itemize}
	\item \textbf{glitch voltage}: The voltage provided to the target during the glitch. This voltage takes values from 1V to 1.6V.
	\item \textbf{glitch length}: The amount of time that the glitch (voltage drop) is applied to the target. In our experiments, this takes up to a small number of clock cycles.
	\item \textbf{glitch delay}: The amount of time between the trigger and the glitch. It should be smaller than the total execution time of the test that runs on the CPU. Similar to the EMFI experiments, the delay used was a random value from 35\% to 65\% of the test's execution time.
\end{itemize}

\begin{figure}[!ht]
	\centering
	\includegraphics[width=0.8\linewidth]{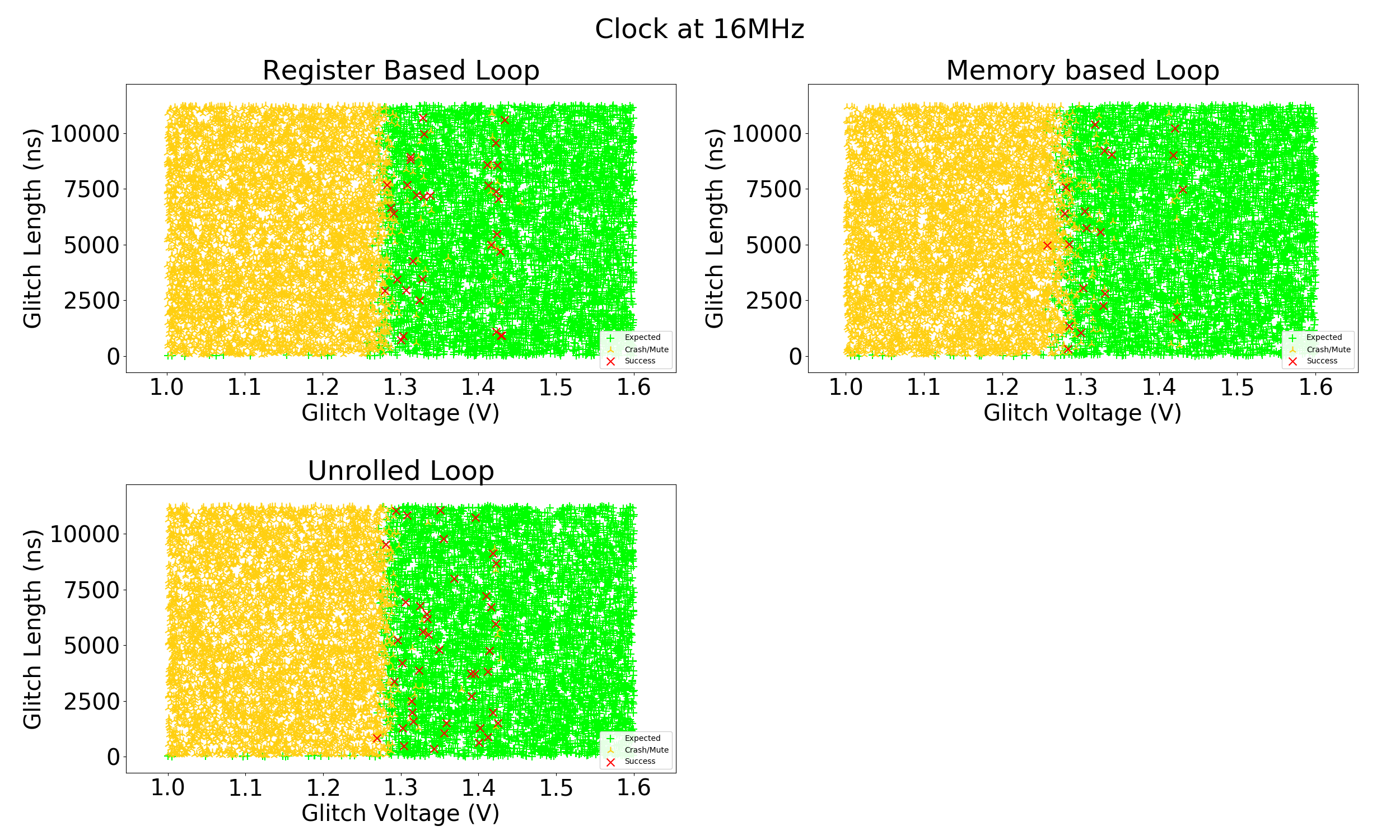}
	\caption{VFI results of all three tests at 16MHz (PLL bypassed).}
	\label{fig:vfi-results-slow}
\end{figure}

\begin{figure}[!ht]
	\centering
	\includegraphics[width=0.8\linewidth]{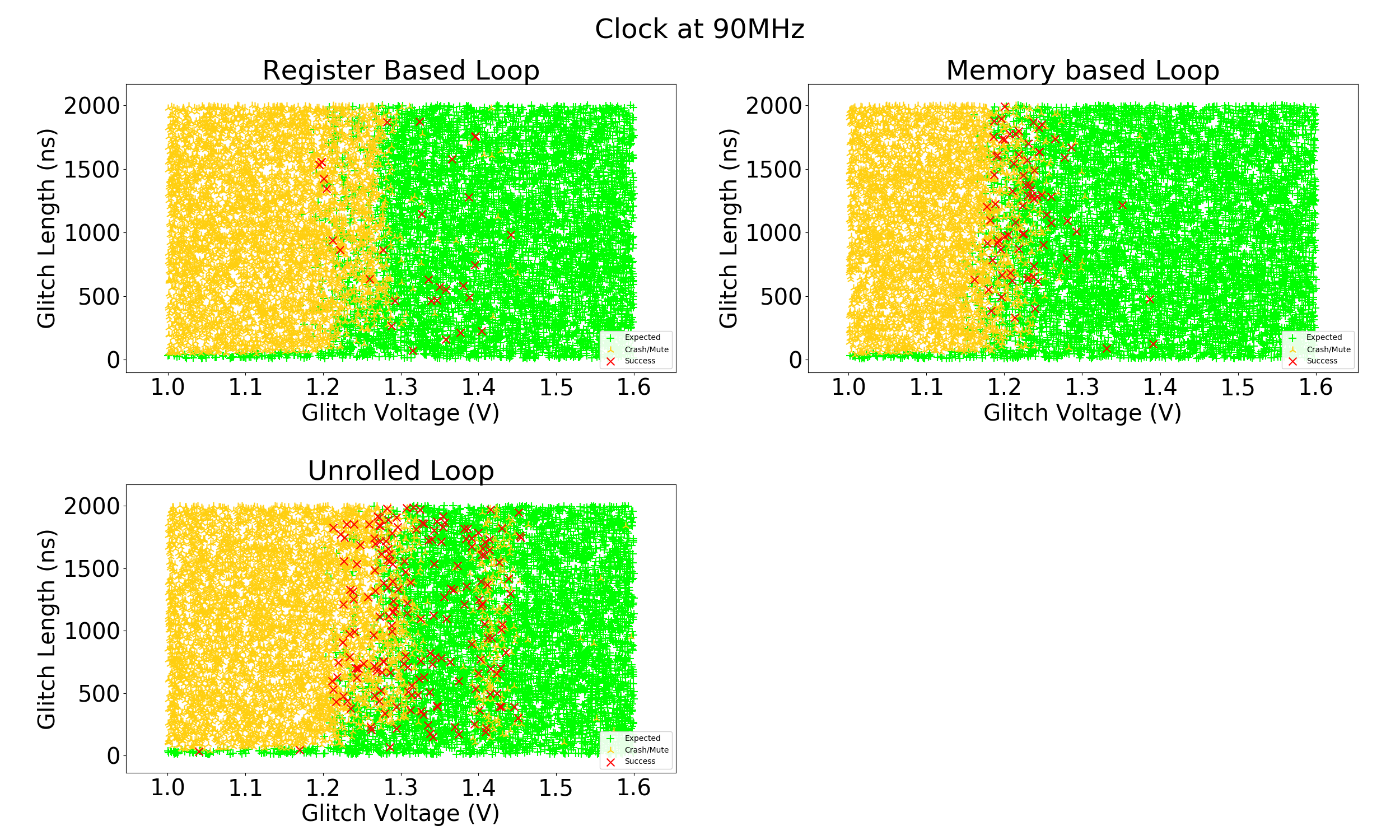}
	\caption{VFI results of all three tests at 90MHz (PLL enabled).}
	\label{fig:vfi-results-medium}
\end{figure}

\begin{figure}[!ht]
	\centering
	\includegraphics[width=0.8\linewidth]{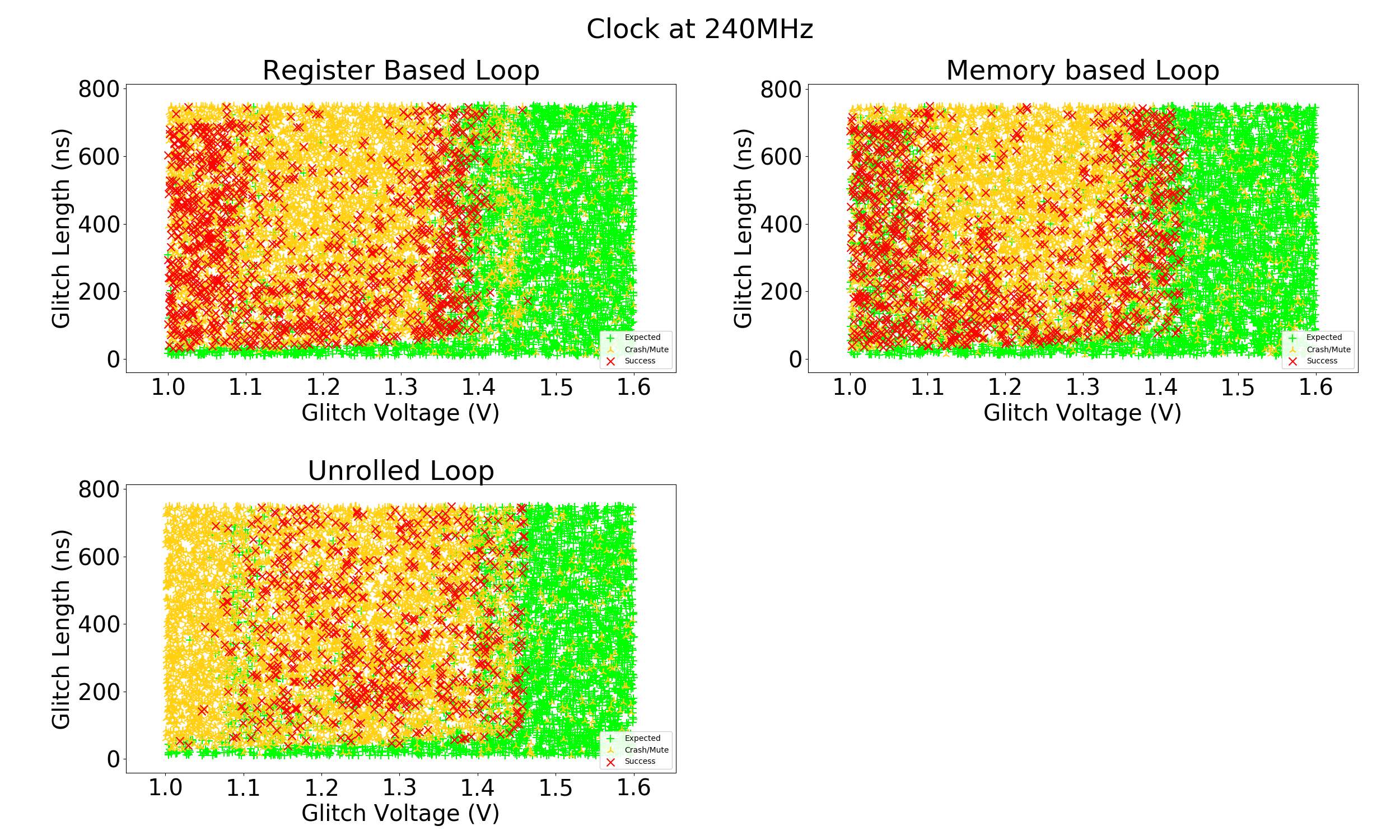}
	\caption{VFI results of all three tests at 240MHz (PLL enabled).}
	\label{fig:vfi-results-fast}
\end{figure}

In our experiments, we varied all three parameters for each attempt. The values for these parameters were chosen randomly from the allowed ranges. Similar to the EMFI, we chose \emph{n} equal to 10000 (0x2710). We show the results from the VFI experiment in~\Cref{fig:vfi-results-slow,fig:vfi-results-medium,fig:vfi-results-fast} for the slow, medium, and fast clock, respectively. The X-axis shows the glitch voltage in volts and the Y-axis shows the glitch length in nanoseconds. Here, the \textit{expected} results, \textit{crashes/mutes}, and \textit{successful} results are marked in \textit{green}, \textit{yellow}, and \textit{red}, respectively (\autoref{sec:res-clas}). The highest frequency that the application could run normally, when no glitch is applied, is 240MHz due to the instabilities that we mentioned in~\autoref{sec:vfi-setup}.

Our experiments show that the largest number of successful glitches appear in the fast clock configuration (240MHz), and the smallest number of successful glitches appear when the circuit operates at 16MHz. In~\cite{zussa2013power}, it was shown that VFI increases signal propagation delays creating timing constraint violations. Such violations become easier when the circuit operates in a higher clock frequency due to the decreased clock period.

In~\autoref{tab:vfi-results-examples}, we show some of the observed successful attempts. Their classification is based on the analysis we presented in~\autoref{sec:res-clas}. When the clock operated at 16MHz, we see that the branch instruction can be skipped in both the register-based loop and the memory-based loop. Here, we got ($t0$, $t1$) = (0x1a5, 0x256b) for the register based loop and ($t0$, $t1$) = (0x1182, 0x158e) and ($t0$, $t1$) = (0x1adb, 0x1adc) for the memory-based loop. In the second case for the memory-based loop, $t0 + t1$ $\neq $ 0x2710, meaning that apart from the branch skipping, one more operation (\texttt{add/sub}) was also manipulated. Instruction manipulation was possible in both the register-based loop (($t0$, $t1$) = (0x25b6, 0)) and the  unrolled loop ($t0$ = 0x26e0). Furthermore, the \texttt{add} instruction was also successfully skipped both in the memory-based loop (($t0$, $t1$) = (0x270f, 0)) and in the unrolled loop ($t0$ = 0x270f).

Similarly, when the clock operated at 90MHz various faults have been observed. The branch instruction was successfully skipped for both the register-based loop (($t0$, $t1$) = (0xeb2, 0x185e)), and the memory-based loop (($t0$, $t1$) = (0xc37, 0x19d9), ($t0$, $t1$) = (0x608, 0x2109), ($t0$, $t1$) = (0xa3d, 0xdeadbeee)). Apart from branch skipping, we also observed memory corruption for the memory-based loop (($t0$, $t1$) = (0xa3d, 0xdeadbee)). Here, after the wrong value was loaded in $t1$, the subtraction was performed and then the branch was skipped. In one case, both the \texttt{sub} and the branch instructions have been skipped (($t0$, $t1$) = (0x608, 0x2109)) because the sum is 0x2711 and $t0$ $\neq 0$. The rest of the experiments (register-based: ($t0$, $t1$) = (0x2700, 0), memory-based: ($t0$, $t1$) = (0x270e, 0), unrolled: $t0$ = 0x2f0f, $t0$ = 0x2711) indicate some kind of instruction manipulation in the constants that were added or subtracted.

When the clock operated at 240MHz, similar results have been observed. In this frequency, there were multiple cases of branch skipping (register-based: ($t0$, $t1$) = (0xb85, 0xdeadbeed), ($t0$, $t1$) = (0x5f0, 0x2120), memory-based: (0xada, 0x1c36), ($t0$, $t1$) = (0xcaa, 0xdeadbeee)). There were also multiple examples of memory corruption for both the register-based loop (($t0$, $t1$) = (0xb85, 0xdeadbeed)) and the memory-based loop (($t0$, $t1$) = (0xdeadce5f, 0), ($t0$, $t1$) = (0xcaa, 0xdeadbeee)). For one of these cases (memory-based: ($t0$, $t1$) = (0xdeadce5f, 0)), the wrong value was loaded to $t0$ and the loop continued normally for 0xdeadce5f - 0xdeadbeef = 0xf70 iterations before it stopped. In the other two cases, the predefined value, i.e., 0xdeadbeef was loaded to $t1$, and the \texttt{sub} instruction decreased $t1$ by 1 (normal execution) or 2 (instruction manipulation), and the branch instruction was skipped in the same loop iteration. In one case for the register-based loop, we got ($t0$, $t1$) = (0x2711, 0). This happened either because one addition was manipulated and a 2 was added instead of 1, or one subtraction was skipped and thus, the loop was executed for 1 more iteration. The rest of the experiments are either instruction manipulations (memory-based: ($t0$, $t1$) = (0x1c0f, 0), unrolled: $t0$ = 0x2720, $t0$ = 0x181c) or addition skipping (unrolled: $t0$ = 0x270f).

\begin{table}[h]
	\centering
	\caption{Output from some of the successful faults in VFI}
	\resizebox{0.95\textwidth}{!}{%
	\begin{tabular}{|l|l|l|l|l|}
	\hline
	Frequency & Test & t0 & t1 & Comment \\ \hline
	\multirow{7}{*}{16MHz} & \multirow{2}{*}{register based loop} & 0x1a5 & 0x256b & instruction skipping (branch) \\ \cline{3-5}
		&  & 0x25b6 & 0x0 & instruction manipulation (add)\\ \cline{2-5}
	 & \multirow{3}{*}{memory based loop} & 0x1182 & 0x158e & instruction skipping (branch) \\ \cline{3-5}
	 &  & 0x270f & 0x0 & instruction skipping (add) \\ \cline{3-5}
		&  & 0x1adb & 0x1adc & instruction manipulation (add) +
		instruction skipping (branch) \\ \cline{2-5}
	 & \multirow{2}{*}{unrolled loop} & 0x270f & - & instruction skipping (add) \\ \cline{3-5}
		&  & 0x26e0 & - & instruction manipulation (add) or instruction
		skipping (add)\\ \hline
	\multirow{9}{*}{90MHz} & \multirow{3}{*}{register based loop} & 0x2700
		& 0x0 & instruction manipulation (add) \\ \cline{3-5}
	 &  & 0xeb2 & 0x185e & instruction skipping (branch) \\ \cline{3-5}
		&  & 0x4b38 & 0x0 & instruction manipulation (add) \\ \cline{2-5}
	 & \multirow{4}{*}{memory based loop} & 0x270e & 0x0 & instruction
		skipping (add) or instruction manipulation (sub or add) \\ \cline{3-5}
	 &  & 0x608 & 0x2109 & instruction skipping (sub + branch) \\ \cline{3-5}
		&  & 0xa3d & 0xdeadbeee & memory corruption + instruction
		skipping (branch) \\ \cline{3-5}
	 &  & 0xc37 & 0x19d9 & instruction skipping (branch) \\ \cline{2-5}
	 & \multirow{2}{*}{unrolled loop} & 0x2f0f & - & instruction
		manipulation (add) \\ \cline{3-5}
		&  & 0x2711 & - & instruction manipulation (add) \\ \hline
	\multirow{10}{*}{240MHz} & \multirow{3}{*}{register based loop} &
		0x2711 & 0x0 & instruction manipulation (add) or instruction
		skipping (sub)\\ \cline{3-5}
		&  & 0xb85 & 0xdeadbeed & memory corruption + instruction
		skipping (branch) \\ \cline{3-5}
		&  & 0x5f0 & 0x2120 & instruction skipping (branch) \\ \cline{2-5}
	 & \multirow{4}{*}{memory based loop} & 0x1c0f & 0x0 & instruction
		manipulation (add or sub) \\ \cline{3-5}
	 &  & 0xada & 0x1c36 & instruction skipping (branch) \\ \cline{3-5}
	 &  & 0xdeadce5f & 0x0 & memory corruption \\ \cline{3-5}
	 &  & 0xcaa & 0xdeadbeee & memory corruption + instruction skipping (branch) \\ \cline{2-5}
	 & \multirow{3}{*}{unrolled loop}
		& 0x2720 & - & instruction manipulation (add) \\ \cline{3-5}
		&  & 0x181c & - & instruction manipulation (add) or instruction skipping (add)\\ \cline{3-5}
	 	&  & 0x270f & - & instruction skipping (add) \\ \hline
	\end{tabular}%
	}
	\label{tab:vfi-results-examples}
\end{table}

\section{Discussion}
\label{sec:disc}

In this section, we present possible explanations for the observed results. The observed results are compatible with earlier analyses of the effects of FI on digital integrated circuits. VFI causes timing constraint violations~\cite{zussa2013power}, which in turn cause computation faults. The timing constraints essentially dictate that the time taken by a circuit to process data must be lower than the clock period of the target for it to function correctly. So, by increasing the data processing time using FI, it is possible to violate the above constraint and induce faults in the computation. As the operating frequency of the target increases, the clock period decreases. Hence, it is relatively easier to violate the setup time constraint, thereby increasing the success rate of VFI.

In EMFI, we did not see only timing faults, but we also observed bit sets, resets, and flips, meaning that our experiments are aligned with the charge-based fault model~\cite{emfi-on-arm-and-riscv} instead of the sampling fault model~\cite{ordas2014evidence}. The sampling fault model~\cite{ordas2014evidence} states that the susceptibility windows of the DFFs are independent of the operating frequency, but the distance of these windows decreases as the clock period gets smaller~\cite{emfi-how-faults-occur}. As a result, one could claim that a glitch injected randomly during the execution of a program has a higher probability of causing a successful fault when the chip operates at a high frequency. However, the fact that we were able to successfully inject faults (not related to the SPI flash) only when the chip operated at 320MHz suggests that the charge-based fault model is more accurate in this case.

In theory, the same success rate can be achieved when the target operates at a lower frequency, e.g., by inducing more powerful voltage glitches or EM pulses. However, the power of glitches cannot arbitrarily be increased in practice without causing the target to reset. This behavior was also observed in our experiments: when the target was running at 240MHz (VFI), we could get many successful faults, where glitch duration was less than 800ns (see~\autoref{fig:vfi-results-fast}). However, we needed to increase the glitch duration for the slower clock speeds, i.e., 2000ns and 12000ns for 90MHz and 16MHz respectively (see~\autoref{fig:vfi-results-medium} and~\autoref{fig:vfi-results-slow}). Such longer glitches inevitably make the target dysfunctional, leading to more resets, as seen from these results. To conclude, when the target operating frequency is low, the success rate decreases due to more resets caused by the increased glitch power. This might also explain the higher success rate when the operating frequency is higher.

\section{Conclusion}
\label{sec:conc}

Many embedded systems in use today are implemented using multi-core SoCs that are complex and host CPUs that run at hundreds of MHz to few GHz. The security of these devices faces different challenges compared to other simple devices like smart cards. In this paper, we investigated the effect of clock frequency on the success rate of VFI and EMFI on such SoCs. To determine the effect of faults more holistically, we developed three test applications that target different components of the SoC. We performed both VFI and EMFI on a RISC-V-based SoC while it was executing our tests. The experimental results showed that the probability of success for fault injection attacks increases as the clock frequency increases. We saw this behavior in both VFI and EMFI. Finally, we provided theoretical justification for the observed results.

\printbibliography

\end{document}